\documentclass[preprint,superscriptaddress]{revtex4-1}
\usepackage{graphicx}
\usepackage{subfigure}
\usepackage{epstopdf}
\usepackage{amsmath}
\usepackage{amssymb}
\usepackage{amsfonts}
\usepackage{mathrsfs}
\usepackage{theorem}
\usepackage{bm}
\usepackage{url}
\usepackage[T1]{fontenc}
\usepackage{csquotes}
\MakeOuterQuote{"}
\usepackage{algorithm}
\usepackage{algorithmicx}
\usepackage{algpseudocode}

\usepackage{dcolumn}
\usepackage{color}
\usepackage{textcomp}
\usepackage[colorlinks=false]{hyperref}

\definecolor{ngreen}{rgb}{0.2,0.6,0.2}

\definecolor{ngold}{rgb}{0.7,0.6,0.2}

\newcommand{\be}{\begin{equation}}
	\newcommand{\ee}{\end{equation}}
\newcommand{\ba}{\begin{align}}
	\newcommand{\ea}{\end{align}}
\def\<{\langle}  
\def\>{\rangle}  

\def\eqref#1{\textup{(\ref{#1})}}  

\newcommand{\cref}[1]{Conjecture~\ref{#1}}
\newcommand{\Cref}[1]{Conjecture~\ref{#1}}

\begin{document}


 \title{Accessing 100~GHz Mechanical Modes in Bulk Crystals at Cryogenic Temperatures}
    \date{\today}
	\begin{abstract}

Sub-terahertz electromechanics offers a promising route to probe mechanical quantum motion at experimentally friendly Kelvin temperatures. Traditionally, high-frequency mechanical resonators rely on advanced microfabrication to shape complex microstructures, while bulk crystals have been largely overlooked due to their large inertia and challenging transduction at such frequencies. Here we show that bulk lithium niobate can host mechanically accessible modes near 100~GHz when coupled via plug-and-play three-dimensional microwave cavities. This approach enables efficient, non-contact excitation of centimeter-scale, milligram-mass vibrational modes across 7.0--110~GHz, with mechanical quality factors up to 30,000 at W band. Furthermore, using a frequency-tunable superconducting niobium cavity at 4~K, we demonstrate strong coupling between a microwave cavity mode and multiple mechanical modes, enabling coherent energy exchange between microwave photons and mechanical phonons with cooperativity up to 16.6 at 110~GHz. These results establish a versatile platform for accessing massive high-frequency mechanical modes and for precision tests of mechanical quantum physics at elevated temperatures.

\end{abstract}
 \author{Boxuan Tian}
	\affiliation{Department of Electrical Engineering, Yale University, New Haven, Connecticut 06520, USA}
 \author{Jiacheng Xie}
	\affiliation{Department of Electrical Engineering, Yale University, New Haven, Connecticut 06520, USA}
    \author{Hong X. Tang}
	\affiliation{Department of Electrical Engineering, Yale University, New Haven, Connecticut 06520, USA}
	\email{hong.tang@yale.edu}
	\maketitle
\emph{Introduction.}-- 
The ability to control and measure the mechanical motion of a bulk crystal at frequencies approaching the sub-terahertz (sub-THz) regime opens a route to exploring quantum phenomena in macroscopic systems. Mechanical resonators operating near 100~GHz possess energy quanta large enough to suppress thermal occupation even at Kelvin-scale temperatures, enabling ground-state preparation with cost-effective cryocoolers and without requiring dilution refrigeration, or involving feedback or side-band cooling  \cite{ref15,ref16, ref17}. Such ultra high-frequency phononic systems are therefore ideally suited for hybrid quantum technologies \cite{ref10,ref12, ref38, ref41,ref30, ref49} to couple with millimeter-wave qubits \cite{ref36} or neutral atoms \cite{ref37}, frequency conversion via optomechanics \cite{ref9,ref13,ref14,ref24}, non-classical mechanical state generation \cite{ref27,ref28,ref29,ref31, ref33,ref39, ref40}, and precision quantum metrology for fundamental physics \cite{ref41,ref42,ref43,ref44,ref45,ref53} at elevated temperatures, while also supporting compact and broadband architectures for classical information processing \cite{ref46}. 

Recent advances in high-frequency (>50~GHz) mechanical resonators based on thin-film lithium niobate (TFLN) \cite{ref1,ref2,ref3,ref4,ref5,ref6,ref52} have pushed acoustic mode frequencies beyond 200~GHz \cite{ref5}. These platforms typically leverage state-of-the-art fabrication tools to create complex micromechanical structures and employ contact electrodes to drive single- or multilayer TFLN structures via the piezoelectric effect \cite{ref1,ref2,ref3,ref5,ref6,ref52}, or to couple mechanical vibrations through optomechanical interactions \cite{ref4}. However, such devices inherently rely on nanoscale features and direct electrode contact for excitation. Owing to the ultrashort phonon wavelength ($\approx 35.7$~nm at 100~GHz) and significant contact-induced loss, the surface defects \cite{ref5} and electrode loading \cite{ref7, ref34, ref35} can substantially limit the achievable mechanical quality factors. Consequently, typical sub-terahertz mechanical resonators in this class exhibit quality factors on the order of hundreds \cite{ref2,ref3,ref4,ref5,ref6,ref52},  and the associated nanosecond phonon lifetimes limit their usefulness for quantum-information applications. A promising route to mitigate these limitations is to employ bulk lithium niobate (bulk LN), whose large volume-to-surface ratio supports low-loss propagation of acoustic waves within the crystal. The use of bulk LN crystal reduces the electromechanical coupling coefficient compared with thin films. Nevertheless,
this drawback can be mitigated by concentrating the microwave field within the crystal using a high-quality microwave cavity\cite{ref9,ref10,ref11}. 

In this work, we develop a plug-and-play, non-contact, piezo-electromechanical coupling scheme that minimizes mechanical damping and eliminates contact-induced loss utilizing a three-dimensional (3D) cavity, which maximizes the overlap between the microwave field and high-order acoustic modes in bulk LN. Mechanical modes with effective vibrational masses of 0.21--5.0~mg and high mechanical quality factors up to 30,000 are observed across the X- to W-bands. Furthermore, by maximizing the microwave cavity quality factor through the use of frequency-tunable niobium superconducting cavities and cooling to 4~K, the system enters the strong-coupling regime \cite{ref20,ref21} with cooperativity $C = 12.3$ and $16.6$ at 85~GHz and 110~GHz, respectively. Together, these results demonstrate a highly efficient and low-loss photon--phonon platform, representing an important step toward realizing sub-terahertz macroscopic quantum mechanics at elevated temperatures.

\begin{figure*}[t]
    \centering
    \includegraphics[width=1\linewidth]{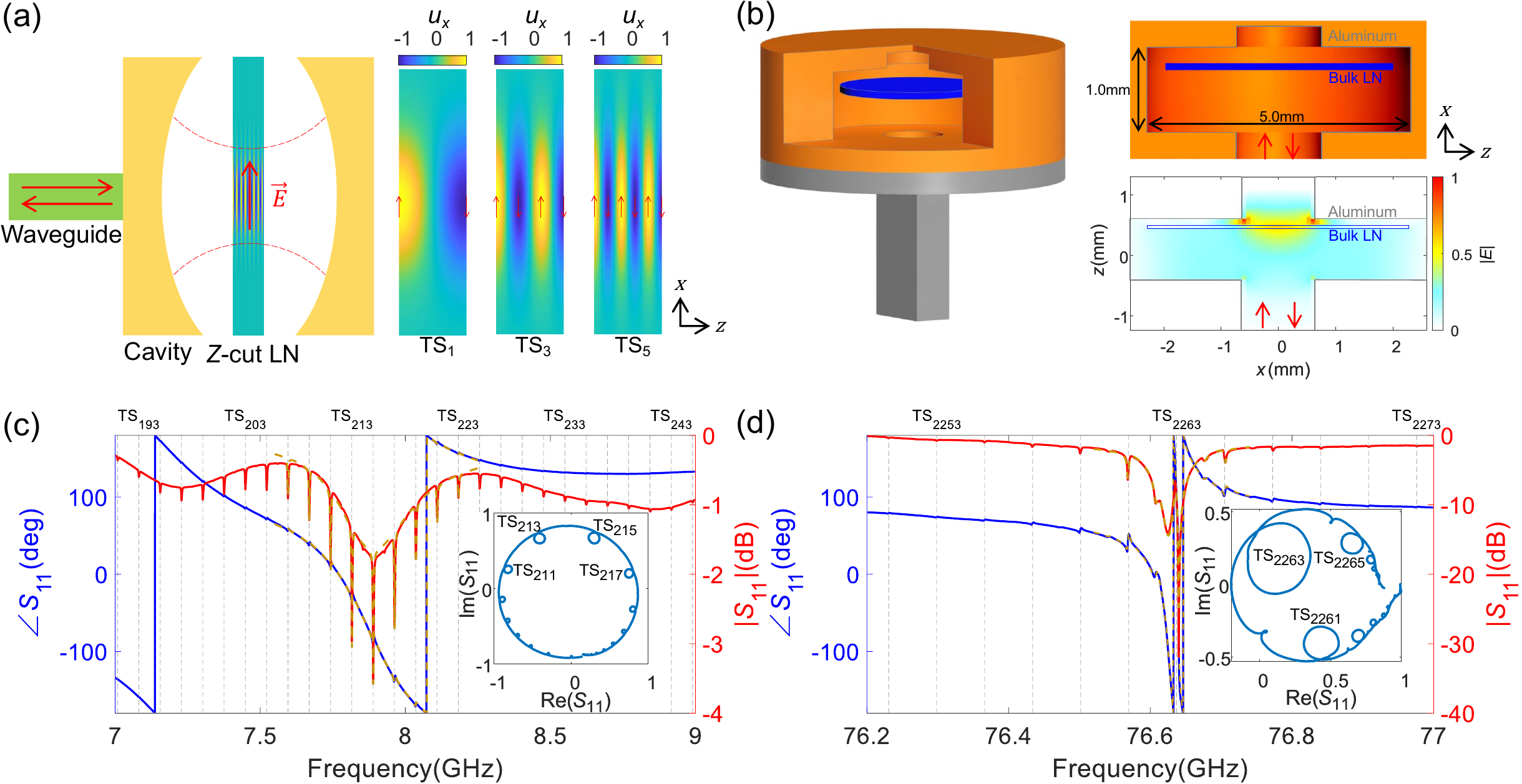}
    \caption{
    Bulk crystal cavity piezo--electromechanical coupling.
    (a) Schematic diagram of the one-port cavity piezo-mechanics system (left) and illustration of the 1st, 3rd and 5th order thickness-shear mechanical modes supported in the bulk LN crystal (right). 
    (b) The model of the W band cavity (left, not to scale), the $x$--$z$ cross-section of the cavity (top right), and the simulated relative electric field of the cavity microwave mode (bottom right). The simulation highlights electric-field confinement in and near a millimeter-size area of the bulk crystal.
    (c) Measured reflection $S_{11}$ spectra and corresponding Smith chart of the X-band cavity-phonon system at room temperature in air.  The equally spaced black dashed lines denote the 191\textsuperscript{st} to 245\textsuperscript{th} TS modes, which generate sharp absorption peaks and clear resonance circles in the S11 spectrum and the Smith chart, respectively.
    (d) Measured reflection $S_{11}$ spectra and corresponding Smith chart of the W-band cavity-phonon system at 6~K. The equally spaced black dashed lines denote the 2251\textsuperscript{st} to 2273\textsuperscript{rd} TS modes. 
    The brown dashed curves in (c) and (d) are the fitting results.
    }
    \label{Fig.1}
\end{figure*}

\emph{Cavity Piezo-electromechanics.—}
The basic principle of the 3D cavity piezo-electromechanics system is illustrated in Fig.~1(a) (left). A $z$-cut bulk LN crystal is placed inside a single-port microwave cavity, where its high-order thickness-shear (TS) mechanical modes couple to the in-plane microwave electric field of a microwave cavity mode via the large piezoelectric coupling element $e_{15}$. By connecting the cavity to a standard waveguide, the microwave reflection spectra (\(S_{11}\)) is measured using a commercial vector network analyzer (VNA). 

For piezoelectric cavity--phonon systems, the interaction between a cavity photon mode and a mechanical phonon mode is proportional to the product of the microwave electric-field amplitude and the mechanical displacement. The total Hamiltonian of the system can therefore be written as
\begin{equation}
H = \hbar\omega_a \hat{a}^\dagger \hat{a}
  + \hbar\omega_b \hat{b}^\dagger \hat{b}
  - \hbar g(\hat{a} + \hat{a}^\dagger)(\hat{b} + \hat{b}^\dagger),
\end{equation}
where $g$ denotes the electromechanical coupling rate, and $\hat{a}\,(\hat{a}^\dagger)$ and $\hat{b}\,(\hat{b}^\dagger)$ 
are the annihilation (creation) operators for the cavity photon and mechanical phonon modes, respectively. Under the rotating-wave approximation and considering the presence of multiple mechanical modes, 
the $S_{11}$ spectra incorporating this interaction can be written as \cite{ref11}
\begin{equation}
S_{11}(\omega)
= 1 - \frac{\Gamma_{a,e}}{
-i\Delta_a + \dfrac{\Gamma_{a,i}+\Gamma_{a,e}}{2}
+ \displaystyle\sum_{n}\frac{g_n^{2}}{-i\Delta_{b,n} + \dfrac{\Gamma_{b,n}}{2}} },
\label{eq:S11_multi_mode}
\end{equation}
where $\Delta_a = \omega - \omega_a$ and $\Delta_{b,n} = \omega - \omega_{b,n}$. $\Gamma_{a,e}$ and $\Gamma_{a,i}$ denote the external and internal loss rates of the cavity, respectively.
$\Gamma_{b,n}$ is the damping rate of the $n$th mechanical mode, and $g_n$ represents the corresponding electromechanical coupling rate.

The thickness-shear modes of the bulk crystal are illustrated schematically in Fig.~1(a) (right). 
For a shear acoustic wave, the TS mode order $n$ and the corresponding frequency $f_n$ satisfy
\begin{equation}
f_n  = n\frac{v}{2L},
\end{equation}
where the shear-wave velocity $v$ remains nearly constant at sub-THz frequencies \cite{ref2} ($v = 3571~\mathrm{m/s}$ at room temperature), and $L$ is the thickness of the bulk crystal. Since the microwave electric field is approximately uniform along the thickness direction (z-axis), only odd-order TS modes couple effectively to the field\cite{ref1,ref2,ref3,ref4,ref5,ref6,ref7,ref11}.

\emph{Design of Cavity-Phonon System.--}
The design of cavity–phonon systems is guided by several key considerations. Maximizing the spatial overlap between the microwave field and the mechanical mode increases the electromechanical coupling rate, which is especially important at high frequencies where larger mechanical linewidths demand stronger coupling. A non-contact geometry further minimizes mechanical loss. To suppress decoherence of ultrashort-wavelength acoustic modes arising from crystal thickness variations, the cavity must also confine the microwave field to a $\sim\mathrm{mm}^2$ area while maintaining a high cavity quality ($Q$) factor.

As shown in Fig.~1(b), we address these challenges using a metallic cavity incorporating a dual circular-aperture structure. The lower aperture defines the coupling interface between the cavity and the waveguide, whereas the upper aperture is positioned 100~\textmu m above the suspended 4.5mm in diameter bulk LN disk. The simulated electric-field distribution at 80~GHz closely resembles that of a conventional electrode configuration \cite{ref2}: the field emanates from one side of the aperture, propagates laterally through the LN wafer, and terminates at the opposite side, forming a tightly confined excitation volume. Thanks to the large dielectric constant of LN (>40) \cite{ref51}, about 70\% of the microwave energy is concentrated within a sub-mm$^3$  volume inside the 50 \textmu m-thick bulk crystal, thereby maximizing electromechanical coupling rate. Meanwhile, since only a small fraction of the microwave energy interacts with the metal surface, this geometry suppresses the metal-induced loss and supports high-$Q$ microwave cavity mode.
\begin{figure*}[t]
    \centering
    \includegraphics[width=1\linewidth]{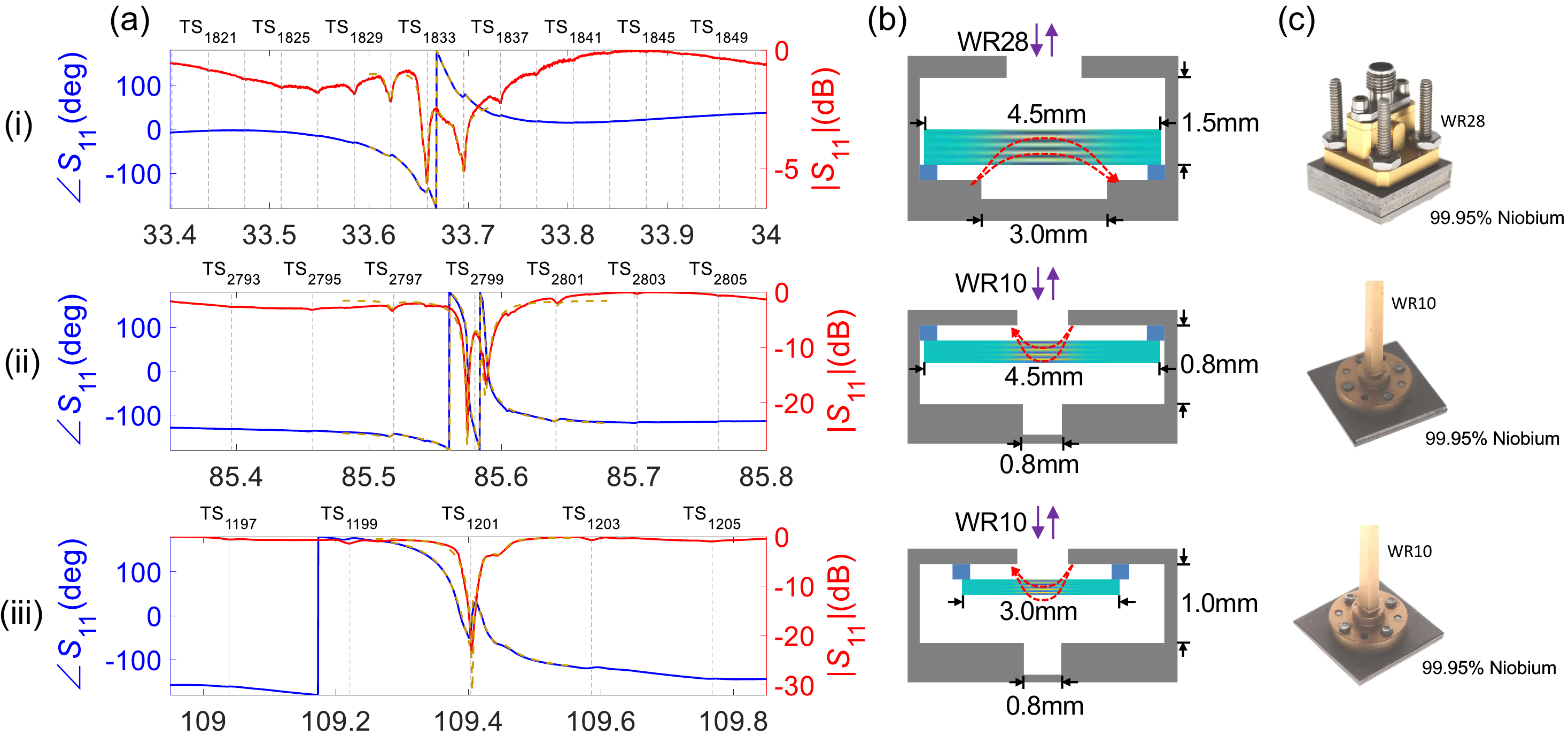}
    \caption{
        Mechanical modes with niobium cavity at cryogenic temperatures (4.2~K). Measured $S_{11}$ spectra (a), schematic illustrations (b) and photographs (c) of the Piezo-electromechanics cavities centered at  (i) 33.67~GHz, (ii) 85.58~GHz, and (iii) 109.40~GHz. The FSRs are 36.7~MHz, 61.2~MHz and 182.2~MHz, respectively. The corresponding bulk LN thicknesses are 97.3~\textmu m, 58.3~\textmu m, and 19.8~\textmu m, respectively, which agree well with the expected values. The cavities are fabricated from pure niobium (superconductor). Refer \cite{ref29, ref41, ref47}, the effective vibration masses are estimated to be 5.0~mg, 0.90~mg, and 0.23~mg, respectively, and the mechanical mode areas are 11~$\rm mm^{2}$, 3.3~$\rm mm^{2}$ and 2.5~$\rm mm^{2}$, respectively (see Supplemental Material for details). For TS$_{1833}$ in (i), TS$_{2799}$ in (ii) and TS$_{1201}$ in (iii), the fitted mechanical quality factors are $7,340\pm310$, $18,900\pm1100$ and $6,260\pm260$, respectively. Due to the FSR estimation uncertainty, the mode order assignments for the three devices carry uncertainties of $\pm4$ mode. The brown dashed curves in (a) are the fitting result. The lower apertures in (b)(ii) and (b)(iii) are prepared for tuning the cavity frequency. 
    }
    \label{Fig.2}
\end{figure*}

\emph{Broadband Mechanical Mode Characterization.--}
To elucidate the basic principles of cavity--phonon coupling, we first investigated a relatively low mechanical-frequency device operating near 8~GHz. The microwave cavity was fabricated from 6061 aluminum alloy, with an inner diameter of 25.4~mm and a height of 19.1~mm. Following Refs.~\cite{ref29, ref41, ref47}, the effective mechanical resonator mass defined by the root-mean-square (RMS) displacement amplitude is estimated to be 2.3~mg. The assembled device was characterized at room temperature in air, with the cavity over-coupled to a WR90 waveguide. The measured $S_{11}$ spectra and the corresponding Smith charts are shown in Fig.~1(c). Within the frequency range of 7--9~GHz, a series of thickness-shear (TS) modes from the 191$^{\textrm{st}}$ to the 245$^{\textrm{th}}$ order are clearly observed, together with the cavity fundamental TE mode at 7.9~GHz, which has a loaded linewidth of $\Gamma_{a,\mathrm{load}}=\Gamma_{a,\mathrm{i}}+\Gamma_{a,\mathrm{e}}=2\pi\times(325.2\pm0.7)$~MHz. From $\mathrm{TS}_{207}$ to $\mathrm{TS}_{223}$, the average fitted mechanical quality factor ($Q_b=f_b/\Gamma_b$) of the 9 odd TS modes is $2,399\pm79$, and the average coupling rate is $2g=2\pi\times(10.42\pm0.10)$~MHz. (See Supplemental Material for the fitting process.)  These mechanical modes exhibit a uniform frequency spacing, with a free spectral range (FSR) of 73.387~MHz, corresponding to a crystal thickness of 48.66~\textmu m, which agrees well with the expected value (50~\textmu m)  and provides strong validation of cavity-phonon coupling scheme.

For high-frequency mechanical resonators, intrinsic phonon--phonon scatterings, including Akhiezer and Landau--Rumer damping, are among dominant room-temperature loss channels~\cite{ref32}. Together with other loss mechanisms, the room-temperature \(fQ\) product record is limited to around \(2.5\times10^{13}\)~\cite{ref2,ref7,ref52}. As a result, the mechanical linewidth may reach hundreds of megahertz in the W-band and surpass the FSR of bulk LN, leading to the difficulty in observing the mechanical modes at the room temperature.

To suppress this fundamental loss, we cooled the device to cryogenic temperatures using a cryocooler. The measured $S_{11}$ spectra of an over-coupled 76.6~GHz cavity made of high-purity aluminum (non-superconducting), together with the corresponding Smith charts, are shown in Fig.~1(d). A series of evenly spaced mechanical modes from TS$_{2251}$ to TS$_{2273}$ are clearly resolved, with FSR of 67.742~MHz and an effective vibrational mass of 1.0~mg. From $\mathrm{TS}_{2261}$ to $\mathrm{TS}_{2265}$, the fitted mechanical quality factors are $30,700\pm1400$, $29,200\pm1400$ and $26,200\pm900$, respectively. The loaded cavity linewidth and the average coupling rate are $\Gamma_{\mathrm{load}}=2\pi\times(72.4\pm0.3)$~MHz and $2g=2\pi\times(13.85\pm0.08)$~MHz, respectively. The resulting cooperativities between the mechanical modes $\mathrm{TS}_{2261}$, $\mathrm{TS}_{2263}$, $\mathrm{TS}_{2265}$ and the cavity are
$C = {4g^2}/({\Gamma_{a,\mathrm{load}}\Gamma_b}) = 1.08\pm0.04$, $0.99\pm0.05$, and $0.91\pm0.03$, respectively, indicating that the coupling approaches the electromagnetically induced transparency (EIT) regime at W band under the over-coupled condition.
 \begin{figure}[t]
    \centering
    \includegraphics[width=1.1\columnwidth]{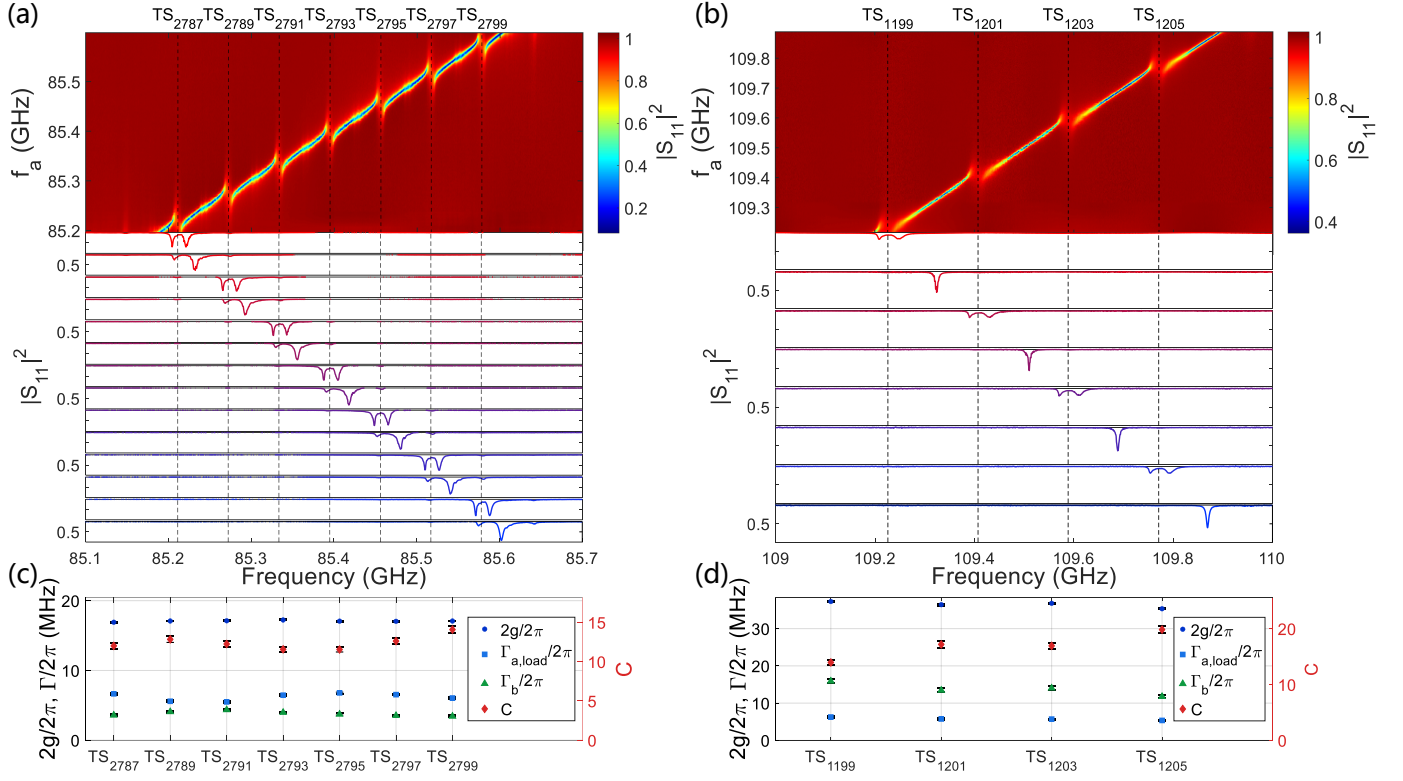}
    \caption{
    Strong coupling between a tunable superconducting cavity and multiple mechanical modes.
    (a,b) Color maps of the normalized reflection spectra and the corresponding linecuts (bottom) for the two cavity--phonon systems near 85.4~GHz and 109.5~GHz, respectively. The bulk LN thickness is 58.3 \textmu m in (a) and 19.8 \textmu m in (b). The devices are cooled down to around 3.7~K in (a) and 4.1~K in (b), respectively. The input power is about -25dBm. The color maps show clear avoided crossings as the cavity frequency is tuned upward and shifts across the selected phonon mode, while the linecuts taken at the phonon resonances and detuned frequencies illustrate mode splitting and hybridization.
    (c,d) Extracted coupling rate $2g$, loaded cavity linewidth $\Gamma_{a,\mathrm{load}}$, mechanical linewidth $\Gamma_b$, and cooperativity $C$ for the corresponding strongly coupled modes in the two devices shown in (a) and (b), respectively.
    }
    \label{Fig.3}
\end{figure}

Furthermore, by varying the cavity dimensions and LN wafer thickness, we observed mechanical modes over a broad frequency range and extended the mechanical resonance frequency to the end of W band. Figs~2a(i)--2a(iii) show the measured $S_{11}$ spectra for over-coupled cavities centered at 33.7~GHz, 85.6~GHz, and 110~GHz, respectively. The clearly resolved mechanical resonances, together with the thickness-dependent FSRs, provide compelling evidence for the observation of genuine high-frequency mechanical modes. At 4.2~K, the superconducting niobium cavity exhibits the highest intrinsic cavity quality factor ($\approx17,000$) among the materials tested. This substantial reduction in cavity loss enables efficient excitation of the TS$_{1833}$, TS$_{2799}$ and TS$_{1201}$ at 33.7~GHz, 85.6~GHz and 110~GHz, respectively.

Figs.~2b(i)--2b(iii) and 2c(i)--2c(iii) illustrate the cavity designs and assembled structures corresponding to these center frequencies. As shown in Figs.~2b(i)--2b(iii), the LN disks are supported near their edges by 100~\textmu m-thick acrylic tape (indicated by the blue squares). The tape is placed approximately 300~\textmu m from the disk edge, with a total contact area of about 1~mm$^2$. This mounting method provides efficient thermalization (see Supplemental Material Section~VII for details) and maintains stable adhesion at cryogenic temperatures. Furthermore, we estimate that the mechanical quality-factor limit imposed by contact loss through the tape is about $2\times10^7$ (see Supplemental Material for details), far above the measured values. Therefore, the contact loss can be safely neglected.

\emph{ Multiple modes strong coupling at W band.--}
We then probe the two devices shown in Figures~2b(ii)--2b(iii) in the under-coupled regime by increasing the lengths of the upper coupling apertures. In addition, we open the lower apertures and extend their lengths to 3~mm. Because these two lower apertures, with diameters of 0.8~mm, have cutoff frequencies higher than the corresponding cavity frequencies, they do not significantly degrade the cavity quality factors. Instead, they allow us to tune the cavity frequency by inserting PTFE dielectric wires into the cavity.

The cavity frequencies are tuned by inserting a 0.031-inch-diameter PTFE wire into the cavities, with its position controlled by a cryogenic piezo stage. Owing to its refractive index of approximately 1.4, the PTFE wire increases the effective mode volume of the cavity, thereby reducing the resonance frequency. The resulting normalized reflection spectra is shown in Figs.~3(a) and 3(b), where tuning ranges of approximately 400~MHz and 600~MHz are demonstrated for the two devices, respectively. As the cavity modes are tuned upward across seven and four mechanical modes, respectively, clear avoided crossings and mode splittings are observed.

For the device shown in Fig.~3(a), fitting the spectra, as summarized in Fig.~3(c), gives an average coupling rate between the cavity mode and the seven mechanical modes of $2g=2\pi\times(17.08\pm0.03)$~MHz. This value exceeds both the average mechanical damping rate $\Gamma_b=2\pi\times(3.81\pm0.05)$~MHz, and the average loaded cavity decay rate $\Gamma_{a,\rm load}=2\pi\times(6.26\pm0.05)$~MHz, confirming that the system has entered the strong-coupling regime with an average cooperativity of $C=12.28\pm0.14$. Similarly, for the device shown in Fig.~3(b), fitting (summarized in Fig.~3(d)) yields an average coupling rate of $2g=2\pi\times(36.24\pm0.10)$~MHz, which is larger than both the average mechanical damping rate $\Gamma_b=2\pi\times(13.52\pm0.23)$~MHz and the average loaded cavity decay rate $\Gamma_{a,\rm load}=2\pi\times(5.56\pm0.06)$~MHz. This results in an average cooperativity of $C=16.60\pm0.27$. The present cavity quality factor is limited by the cooling capacity of the cryocooler, and further improvements are expected if the operating temperature can be reduced to the 1~K regime.

\emph{Polarization degeneracy of the mechanical modes.--}
The high spectral resolving power of our cavity photon--phonon system enables direct investigation of the polarization states of mechanical modes. As illustrated in Fig.~1(d), several devices exhibit additional features in the $S_{11}$ spectra that are not associated with the primary mode family. Further analysis identifies these features as arising from polarization degeneracy. As shown in Fig.~4, rotating the cavity by $90^\circ$ switches the dominant coupling from the set of horizontally polarized (H) modes to the set of vertically polarized (V) modes, revealing a fractional FSR difference of $0.558\%$. This small FSR mismatch likely arises from a slight wafer miscut or thickness nonuniformity, which breaks the in-plane $x$--$y$ symmetry of the $z$-cut bulk LN and thereby establishes fast and slow acoustic axes. Such polarization splitting introduces an additional degree of freedom into the cavity--phonon system, enabling both linearly and circularly polarized mechanical motion in the $x$--$y$ plane through frequency modulation.

\begin{figure}[t]
    \centering
    \includegraphics[width=0.75\columnwidth]{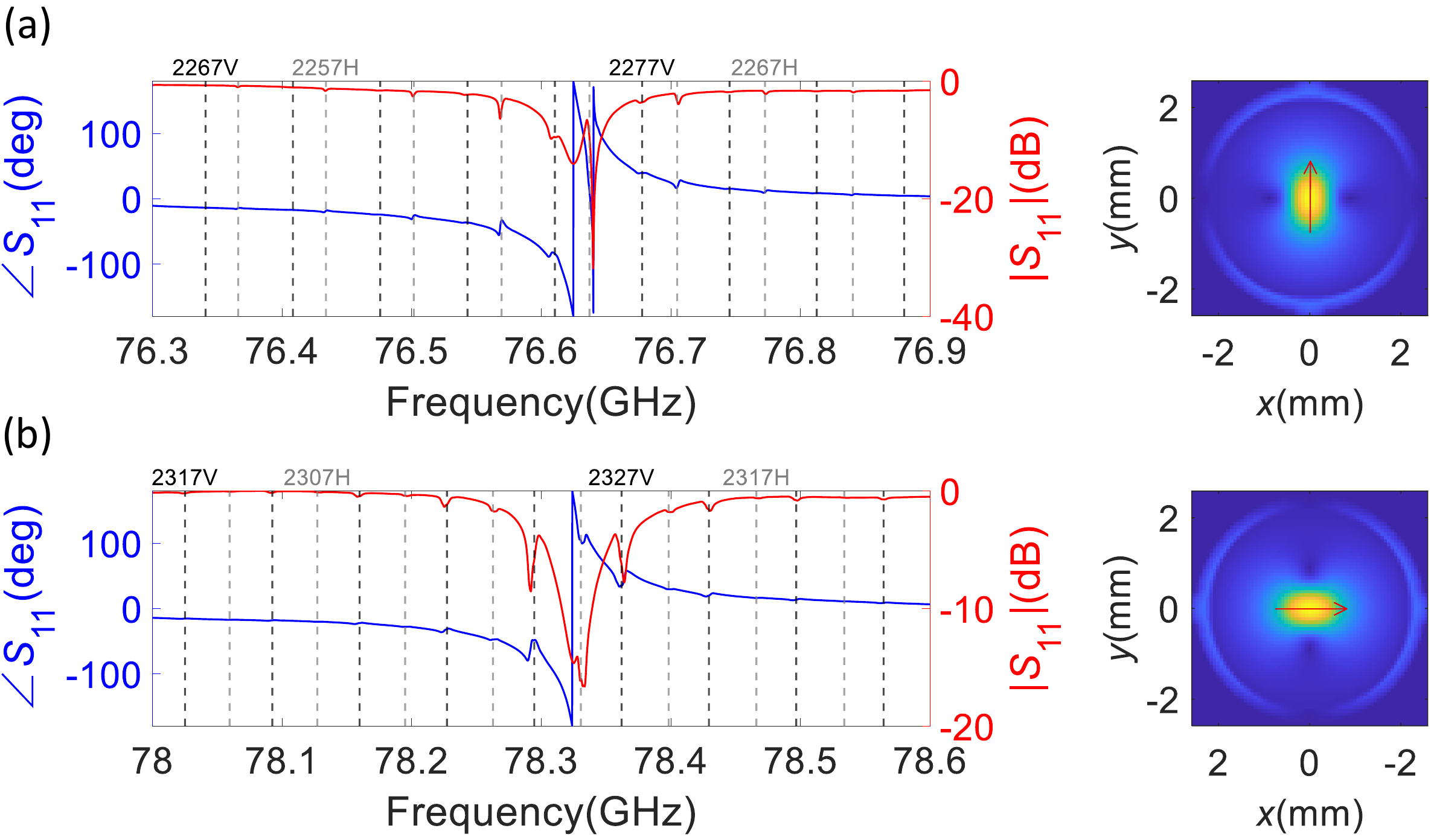}
    \caption{
        Polarization splitting of mechanical modes. 
        (a,b) $S_{11}$ spectra of the same aluminum cavity-phonon system before and after a 90° rotation, respectively, together with simulated electric-field distributions on the bulk LN. 
        The gray and black dashed lines denote two sets of polarized TS modes: the H-polarized and V-polarized modes, respectively. Before rotation, the cavity predominantly excites the H-polarized modes while only weakly coupling to the V-polarized ones; after rotation, the situation is reversed. These two mode families are non-degenerate due to a $0.558\%$ FSR difference, where FSR$_\mathrm{H}$ = 67.742~MHz and FSR$_\mathrm{V}$ = 67.364~MHz.
    }
    \label{Fig.4}
\end{figure}

\begin{figure}[t]
\centering
\includegraphics[width=0.8\columnwidth]{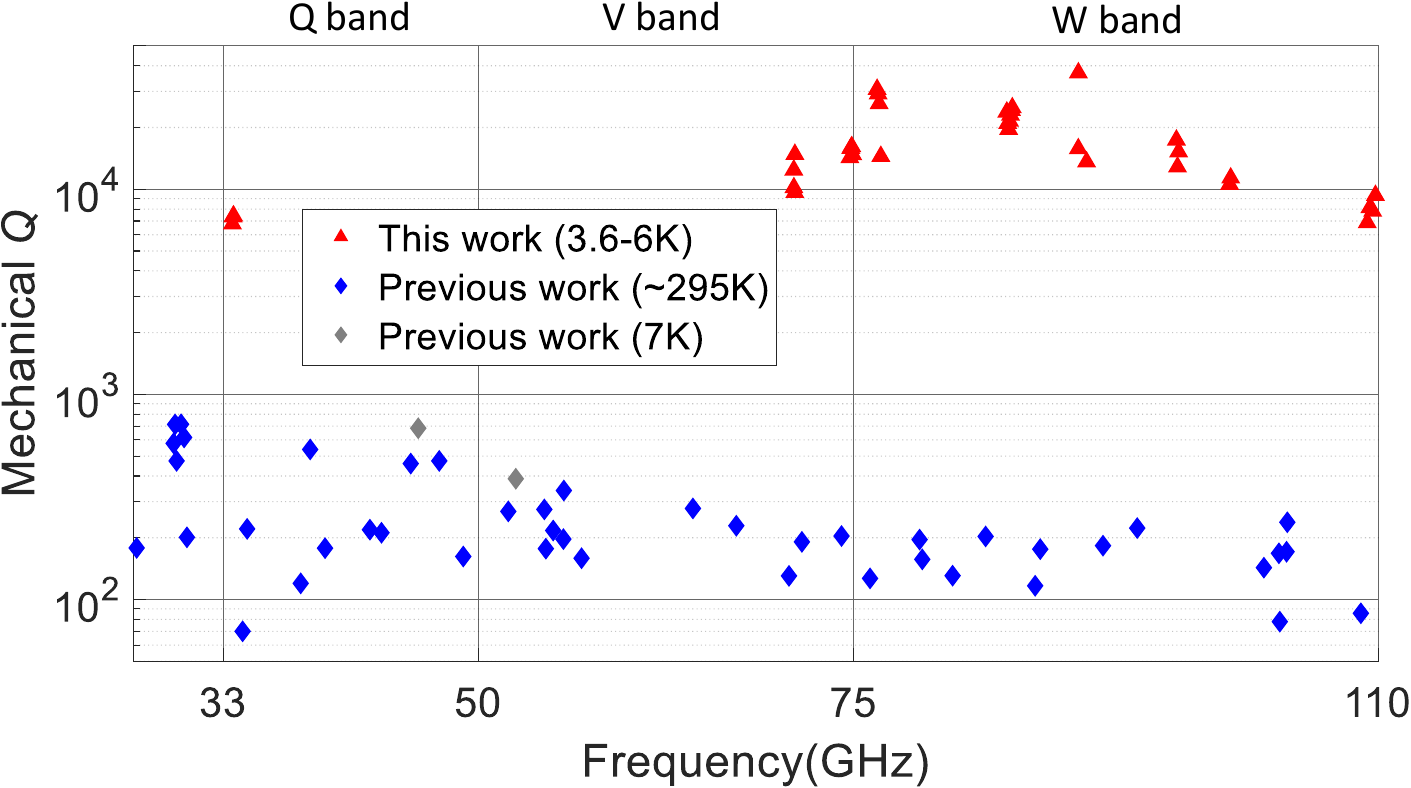}
\caption{Mechanical quality factors $Q$ measured in this work (red triangles), spanning 33--110~GHz. Among the modes from 9 devices with LN thicknesses ranging from 40 to 60~\textmu m, the mean quality factor is $18{,}500$, with a device-to-device standard deviation of $6{,}500$. The highest $fQ$ product observed in this work is $(3.326\pm0.022)\times10^{15}$~Hz at 90.0~GHz (see Supplemental Material for details). The results are compared with a broadband survey of previously reported mm-wave ($>27$~GHz) piezo-electromechanical resonators. Literature data measured at room temperature are shown as blue diamonds~\cite{ref8,ref1,ref2,ref5,ref6,ref7}, while cryogenic data are shown as gray diamonds~\cite{ref52}. This comparison highlights the nearly two-order-of-magnitude enhancement in $Q$ achieved in this work at W band.}
\label{Fig.5}
\end{figure}

\emph{Discussion.--}
In conclusion, we demonstrate a sub-THz bulk-crystal cavity piezo-electromechanical platform that operates in the strong-coupling regime while maintaining low dissipation. As summarized in Fig.~5, eliminating electrode-contact loss, increasing the volume-to-surface ratio of the mechanical resonator, suppressing thermal phonon scattering at cryogenic temperatures, and confining the electric-field energy to a sub-mm$^3$ volume in this work collectively yield nearly two orders of magnitude improvement in mechanical quality factor compared with previous sub-THz piezo-electromechanical devices. These results highlight the intrinsic advantages of bulk LN for realizing ultrahigh-frequency, low-loss phononic systems.

Based on the Bose--Einstein distribution, a 110~GHz mechanical resonator at 4.1~K is expected to have a thermal phonon occupancy of only $0.38$, indicating that our system may be already near its quantum ground state using only a cost-effective cryocooler, without requiring any additional cooling or sideband operations. A strict quantum-noise--limited characterization will, however, require sub-THz quantum-limited microwave amplifiers, which may become available in the near future given emerging developments such as the K band Josephson parametric amplifier \cite{ref48}. For a ground-state harmonic oscillator with a 0.23~mg effective mass and 110~GHz frequency, the zero-point position uncertainty is theoretically estimated to be $\Delta x=\sqrt{\hbar/(4\pi mf)}\approx1.8\times10^{-20}~\mathrm{m}$. Such extraordinarily small displacements may open new opportunities in precision quantum metrology, enabling tests of fundamental physics in previously inaccessible spatial regimes.

Altogether, our work demonstrates a scalable, low-loss bulk crystal cavity electromechanical platform operating up to  sub-THz frequencies, opening pathways toward the exploration of macroscopic quantum motion, sub-THz hybrid quantum technologies, and precision metrology at elevated temperatures.

\vspace{8mm}\noindent{\bf Acknowledgements}\\
This project is supported by the Air Force Office of Sponsored Research (AFOSR MURI FA9550-23-1-0338) and in part by the Defense Advanced Research Projects Agency (DARPA OPTIM HR00112320023).  The authors would like to thank Nicholas Bernardo for assistance in machining the cavities used in this work.

\bibliographystyle{apsrev4-1}
\bibliography{main}

\end{document}


\setcounter{figure}{0}
\renewcommand{\thefigure}{\arabic{figure}}

\begin{center}
{\Large\bfseries Accessing 100~GHz Mechanical Modes in Bulk Crystals at Cryogenic Temperatures: Supplemental Material\par}
\end{center}

\vspace{0.5em}

In the Supplemental Material, we discuss the following topics:

I. Cavity-phonon coupling \(g\) estimation.

II. Fitting method and uncertainty estimation.

III. Transverse mechanical mode analysis.

IV. Estimation of the contact loss.

V. Cavity simulation and mechanical mode area estimation.

VI. Time-domain analysis through Fourier transform.

VII. Power- and Temperature- dependence of the cavity loss.

VIII. Dielectric loading and cavity intrinsic loss analysis.

IX. Experimental setup.

    \section{ Cavity-phonon coupling \(g\) estimation.}

Following the treatment in the Supplemental Material of Ref.~\cite{ref11a}, the piezo-electromechanical
coupling rate for an odd thickness-shear mode can be estimated as
\begin{equation}
    g_{2n+1}
    =
    \frac{e_{\rm eff}\sqrt{2}}
    {\sqrt{\epsilon_0\epsilon_{\rm LN}\rho_{\rm LN}}}
    \frac{\eta}{\sqrt{t_{\rm LN}t_e}} .
\end{equation}
Here \(e_{\rm eff}\) is the effective piezoelectric coefficient for the coupling
between the in-plane microwave electric field and the thickness-shear strain,
\(\epsilon_{\rm LN}\) is the relative dielectric constant of lithium niobate for the relevant
microwave polarization, \(\rho_{\rm LN}\) is the density of lithium niobate,
\(t_{\rm LN}\) is the LN thickness, \(t_e\) is the effective microwave mode depth,
and
\begin{equation}
    \eta
    =
    \frac{A_{\rm overlap}}{\sqrt{A_m A_e}}
\end{equation}
is the dimensionless lateral overlap factor between the microwave and acoustic
modes. Here \(A_{\rm overlap}\) is the lateral overlap area, \(A_m\) is the
effective acoustic mode area, and \(A_e\) is the effective microwave mode area.

The effective microwave mode depth \(t_e\) can be expressed using the microwave
energy concentration ratio in LN. For a nearly uniform microwave electric field
inside the LN,
\begin{equation}
    p_{\rm LN}
    =
    \frac{U_{\rm LN}}{U_{\rm tot}}
    =
    \frac{t_{\rm LN}}{t_e}.
\end{equation}
Therefore,
 $t_e
    =
    \frac{t_{\rm LN}}{p_{\rm LN}} .$
Substituting this relation into the coupling expression gives
\begin{equation}
    2g_{2n+1}
    \approx
    \frac{2\sqrt{2}e_{\rm eff}}
    {\sqrt{\epsilon_0\epsilon_{\rm LN}\rho_{\rm LN}}}
    \eta
    \frac{\sqrt{p_{\rm LN}}}{t_{\rm LN}} .
\end{equation}
For z-cut shear mode in LN, we take
    $e_{\rm eff} = e_{15}$.
Using room-temperature material parameters for LN\cite{ref56},
the coupling rate expression becomes
\begin{equation}
    2g_{2n+1}
    \approx
    2\pi\times
    \left(
    1.28\times10^3~
    \eta
    \frac{\sqrt{p_{\rm LN}}}{t_{\rm LN}}
    \right)~{\rm Hz}.
\end{equation}
where \(t_{\rm LN}\) should be expressed in meters.
We compare the measured coupling rates of the two W-band devices shown in
Figs.~3 of the Main Text with the theoretical estimates.

For the 85.4~GHz device, the LN thickness is 58.3~\textmu m, and the
measured coupling rate is $2g_{\rm measured}=$\(2\pi\times17.08~{\rm MHz}\). Based on the microwave
simulation, we estimate \(p_{\rm LN}\approx0.8\). The theoretical coupling rate is
therefore $2g_{\rm th}
    \approx
    \eta \times 2\pi\times19.7~{\rm MHz}.$

Similarly, for the 109.4~GHz device, the LN thickness is 19.8~\textmu m, and the
measured coupling rate is $2g_{\rm measured}=$\(2\pi\times36.24~{\rm MHz}\). Using
\(p_{\rm LN}\approx0.45\), the theoretical coupling rate is $2g_{\rm th}
    \approx
    \eta \times 2\pi\times43.4~{\rm MHz}.$
 
Assuming \(\eta=1\), the theoretical coupling rates approximately agree with
the experimental values, exceeding the measured values by about \(15.3\%\) and
\(19.8\%\), respectively. This difference
can be attributed to uncertainty in the simulated microwave energy participation, parameter change at the cryogenic temperature, and the nonzero angle between the mechanical displacement direction and the microwave electric field direction caused by polarization degeneracy.

\begin{figure}[t]
\centering
\includegraphics[width=1\columnwidth]{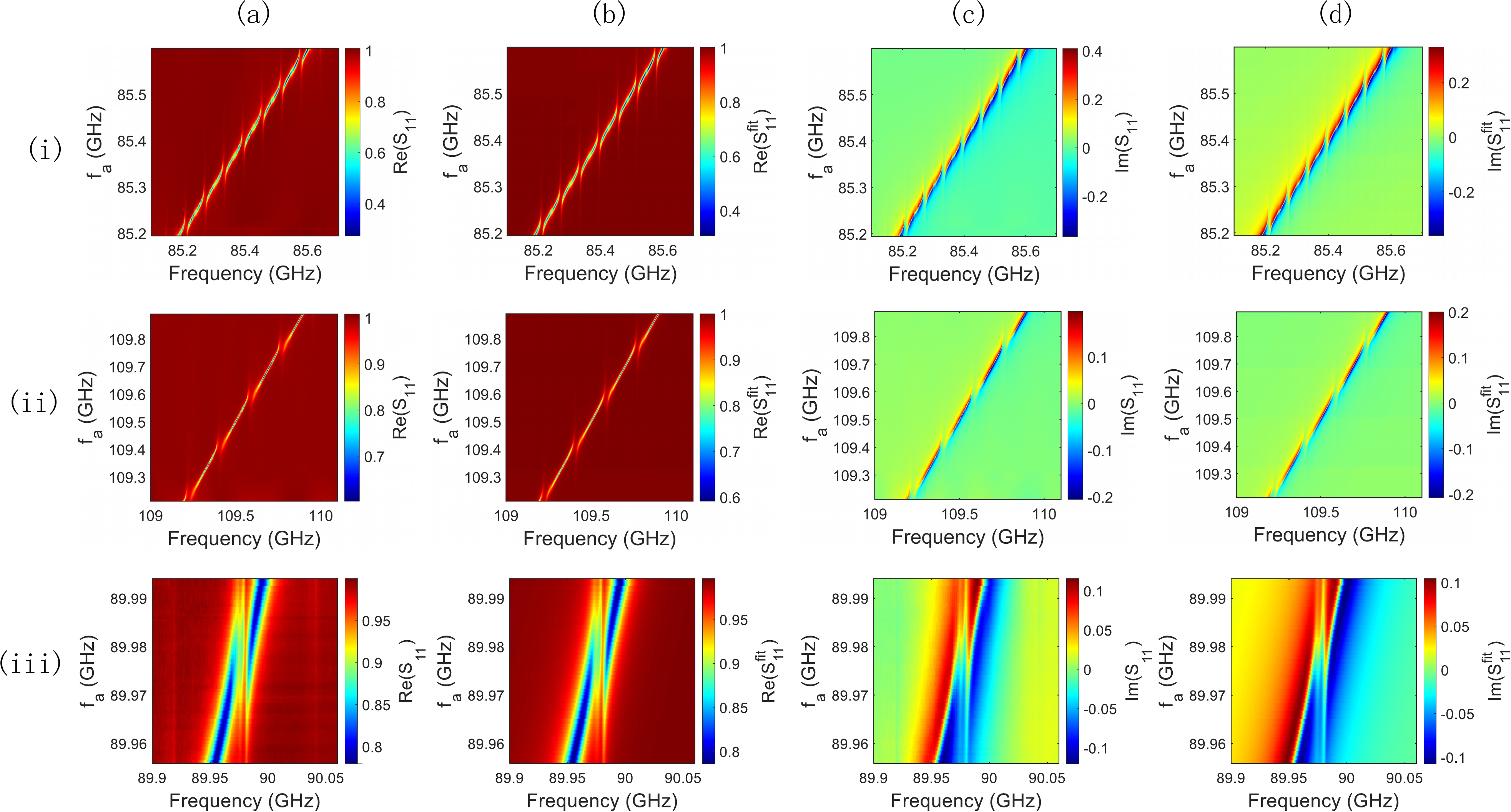}
\caption{(a)--(d) show the measured real part, fitted real part, measured imaginary part, and fitted imaginary part of the reflection spectra, respectively. (i,ii) Measured and fitted complex $S_{11}$ spectra for the two devices presented in Fig.~3 of the Main Text. (iii) Measured and fitted spectra of the device exhibiting the highest mechanical quality factor observed in this work. Two closely spaced mechanical modes are resolved at 89.9746 and 89.9800~GHz, respectively. The corresponding mechanical quality factors are $15{,}770\pm120$ and $36{,}960\pm240$ for the left and right modes, respectively. The mechanical modes near them with $\rm FSR\approx61~$MHz can also be observed at $89.919~$GHz and $90.041~$GHz in the measured spectra.}
\label{Fig.5}
\end{figure}

\section{Fitting Method and Uncertainty Estimation}

The complex $S_{11}$ spectra are fitted using a one-port cavity--phonon input-output model. The fitted complex reflection coefficient is written as
\begin{equation}
S_{\mathrm{fit}}(f)=
Ae^{if\tau}
\left[
1-
\frac{\Gamma_{a,e} e^{i\Phi}}
{
-i(f-f_a)
+
\Gamma_{a,\mathrm{load}}/2
+
\displaystyle\sum_m
\frac{g_m^2}
{
-i(f-f_{b,m})+\Gamma_{b,m}/2
}
}
\right],
\label{eq:Sfit_complex}
\end{equation}
where $f$ is the probe frequency, $\tau$ is the electrical delay, $f_a$ is the microwave cavity resonance frequency, and $f_{b,m}$ is the resonance frequency of the $m$-th mechanical mode. $\Gamma_{a,e}$ is the cavity external coupling rate, $\Gamma_{a,\mathrm{load}}$ is the loaded cavity linewidth, and $\Gamma_{b,m}$ and $g_m$ are the mechanical linewidth and coupling rate of the $m$-th mechanical mode, respectively. The complex coefficient $A$ accounts for attenuation and phase shifts in the transmission line after calibration. The phase $\Phi$  quantifies the impedance mismatch\cite{ref54,ref55}.

The measured data are fitted in the complex plane. The fitting parameters are obtained by minimizing the sum of squared complex residuals within the fitting window,
\begin{equation}
\chi^2=
\sum_i
\left|
S_{\mathrm{fit}}(f_i)-
S_{\mathrm{meas}}(f_i)
\right|^2,
\label{eq:least_square}
\end{equation}
where $S_{\mathrm{meas}}(f_i)$ is the measured complex $S_{11}$ at the $i$-th frequency point. For two-dimensional color maps, the index $i$ runs over all probe-frequency points and cavity-frequency settings.

Parameter uncertainties are estimated by calculating the covariance matrix from the residuals between the fitted model and the measured data using the heteroskedasticity-robust covariance estimator (refer\cite{ref58}). All uncertainties and error bars reported in this work correspond to 2 times the standard error.

As supplementary validation, we show the real and imaginary parts of the measured and fitted complex $S_{11}$ spectra for three devices in Fig.~1. The first and second devices are those presented in Fig.~3 of the Main Text, while the third device exhibits the highest mechanical quality factor observed in this work.

\section{Transverse mechanical mode analysis}

Here we analyze the transverse distribution of the mechanical motion. The microwave electric field is assumed to be uniform along the thickness direction,
\begin{equation}
E_x(x,y,z)\equiv\mathcal E(x,y).
\end{equation}
Here we take the time dependence as \(e^{-i\omega t}\). We decompose the displacement field into a leading semi-one-dimensional solution and a correction term,
\begin{equation}
\mathbf u=\mathbf u_{\rm 1D}+\mathbf u_c .
\end{equation}
The leading semi-1D solution is obtained by assuming the lateral overlap is unity,
\begin{equation}
u_{\mathrm{1D},x}=\mathcal E(x,y)U(z),\qquad
u_{\mathrm{1D},y}=0,\qquad
u_{\mathrm{1D},z}=0 .
\end{equation}
Here \(U(z)\) is the thickness displacement response. To leading order, it satisfies
\begin{equation}
\tilde c_{44}U''+\rho\omega^2 U=0,
\end{equation}
where $\tilde c_{44}$ is the effective complex shear elastic stiffness of LN, and $\rho$ is the mass density. With the dominant free-surface boundary condition
\begin{equation}
\tilde c_{44}\partial_z u_{\mathrm{1D},x}
-
e_{15}\mathcal E(x,y)
=0,
\qquad
z=\pm\frac{l}{2}.
\end{equation}
The solution is
\begin{equation}
U(z)
=
\frac{e_{15}}{\tilde c_{44}}z
+
\sum_{n=1,3,5,\cdots}
\frac{
4e_{15}\sin(n\pi/2)
}{
\rho l
\left(
\tilde\omega_n^2-\omega^2
\right)
}
\sin\left(\frac{n\pi z}{l}\right),
\end{equation}
where
$\tilde\omega_n
=
\frac{n\pi}{l}
\sqrt{\frac{\tilde c_{44}}{\rho}},$ and $
\tilde\omega_n^2-\omega^2
\approx
\omega_n
\left[
2(\omega_n-\omega)-i\frac{\omega_n}{Q_b}
\right]$.

The correction field is written as
\begin{equation}
\mathbf u_c=(u_{c,x},u_{c,y},u_{c,z})^T .
\end{equation}
It satisfies the 3D elastic equations driven by the residual terms left by the semi-1D solution,
\begin{align}
-\rho\omega^2 u_{c,x}
&=
\partial_x T_1^c+\partial_y T_6^c+\partial_z T_5^c+R_x,
\\
-\rho\omega^2 u_{c,y}
&=
\partial_x T_6^c+\partial_y T_2^c+\partial_z T_4^c+R_y,
\\
-\rho\omega^2 u_{c,z}
&=
\partial_x T_5^c+\partial_y T_4^c+\partial_z T_3^c+R_z .
\end{align}
Here \(T_i^c\) are the stresses generated by \(\mathbf u_c\), and the residual body sources are
\begin{align}
R_x
&=
\tilde c_{11}U\partial_x^2\mathcal E
+
\tilde c_{66}U\partial_y^2\mathcal E
+
2\tilde c_{14}U'\partial_y\mathcal E,
\\
R_y
&=
(\tilde c_{12}+\tilde c_{66})U\partial_x\partial_y\mathcal E
+
2\tilde c_{14}U'\partial_x\mathcal E,
\\
R_z
&=
(\tilde c_{44}+\tilde c_{13})U'\partial_x\mathcal E
+
2\tilde c_{14}U\partial_x\partial_y\mathcal E
-
e_{15}\partial_x\mathcal E .
\end{align}

We analyze these equations perturbatively. Here we briefly summarize the calculation results. The correction at each order is governed by the characteristic transverse ratio
\begin{equation}
\epsilon_\perp
=
\frac{\lambda_b}{L_{\mathcal E}}
=
\lambda_b
\left[
\frac{
\displaystyle\int d^2r\,|\nabla_\perp\mathcal E(x,y)|^2
}{
\displaystyle\int d^2r\,|\mathcal E(x,y)|^2
}
\right]^{1/2}.
\end{equation}
From the microwave simulation, \(\epsilon_\perp\approx1.0\times10^{-4}\) for the 85.4~GHz device.

The largest correction comes from the terms in \(R_x\) and \(R_y\) that contain only one lateral derivative of \(\mathcal E\),
\begin{align}
R_x^{(1)}
&=
2\tilde c_{14}U'\partial_y\mathcal E,
&
R_y^{(1)}
&=
2\tilde c_{14}U'\partial_x\mathcal E .
\end{align}
We denote the corresponding correction field by \(\mathbf u_c^{(1)}\). To leading order,
\begin{equation}
\mathbf u_{c}^{(1)}
=
\begin{pmatrix}
u_{c,x}^{(1)}\\
u_{c,y}^{(1)}
\end{pmatrix}
=
\sum_{n=1,3,5,\cdots}
\sum_{m=2,4,6,\cdots}
\frac{
32e_{15}\tilde c_{14} n^2
}{
\rho^2 l^2
(n^2-m^2)
\left(
\tilde\omega_m^2-\omega^2
\right)
\left(
\tilde\omega_n^2-\omega^2
\right)
}
\begin{pmatrix}
\partial_y\mathcal E\\
\partial_x\mathcal E
\end{pmatrix}.
\end{equation}
The relative correction amplitude is approximately
\begin{equation}
\frac{
\left\langle |\mathbf u_{c}^{(1)}|^2 \right\rangle^{1/2}
}{
\left\langle |\mathbf u_{\rm 1D}|^2 \right\rangle^{1/2}
}
\approx
\frac{\sqrt{2}}{\pi}
\left|
\frac{c_{14}}{c_{44}}
\right|
n\epsilon_\perp
\approx
6\times10^{-3},
\end{equation}
where \(n=2793\) for a 58.3~\textmu m-thick LN wafer at 85.4~GHz. Thus, the first-order correction amplitude is about 160 times smaller than the leading semi-1D solution.

The other residual terms are also calculated. The correction from the residual boundary conditions is about 3,700 times smaller than the semi-1D solution, while the correction from body-source terms containing two lateral derivatives of \(\mathcal E\) is about 33,000 times smaller.

Therefore, the transverse mechanical displacement is well approximated by the semi-1D solution, and its lateral amplitude follows the local microwave electric-field amplitude.

\section{Estimation of the Contact Loss}

Based on the transverse mode analysis above, the dominant TS displacement can be well approximated by the semi-1D solution
\begin{equation}
\mathbf u_{\rm 1D}(x,y,z)
=
\begin{pmatrix}
\mathcal E(x,y)U(z)\\
0\\
0
\end{pmatrix},
\end{equation}
where \(U(z)\) is the thickness shear response, and the transverse distribution of the acoustic energy is mainly determined by the local microwave field amplitude $\mathcal E(x,y)$.

We define the transverse acoustic energy density as
\begin{equation}
W(x,y)
=\frac{\rho \omega^2}{2}
\int_{-l/2}^{l/2} dz\, |\mathbf{u_{\rm 1D}}(x,y,z)|^2.
\end{equation}
When assuming that the tape will absorb all the incident acoustic wave, the contact induced energy decay rate can then be written as
\begin{equation}
\Gamma_{\rm contact}
=
\frac{v_s}{2l}
\frac{
\displaystyle
\int_{A_{\rm contact}} d^2r\, W(x,y)
}{
\displaystyle
\int_{A_{\rm wafer}} d^2r\, W(x,y)
}.
\end{equation}
Here \(v_s\) is the shear acoustic velocity and \(l\) is the wafer thickness.  Within the semi-1D approximation,
$W(x,y)
\propto
|\mathcal E(x,y)|^2 .$
Thus the contact loss rate can be estimated through the transverse electric field distribution.

For the 85.4 GHz device shown in Fig.~3 of the Main Text, we use $l=58.3$~\textmu m, \(\lambda_b=41.8~{\rm nm}\), and \(v_s=3.571\times10^3~{\rm m\,s^{-1}}\). The wafer diameter is 4.5~mm, and the contact area is in the edge of the wafer within 300 \textmu m distance and the total contact area is about $1$~mm$^2$. From the simulated microwave field distribution, the acoustic energy contained in the contact region is approximately
\begin{equation}
\frac{
\displaystyle
\int_{A_{\rm contact}} d^2r\, W(x,y)
}{
\displaystyle
\int_{A_{\rm wafer}} d^2r\, W(x,y)
}
\approx
\frac{1}{1100}.
\end{equation}
The corresponding contact loss rate is therefore
$\Gamma_{\rm contact}
\approx
3\times10^4~{\rm s^{-1}}.$
And the corresponding contact loss limited quality factor is
$Q_{\rm contact}
=
\frac{2\pi f_b}{\Gamma_{\rm contact}}
\approx
2\times10^7.$
 Since the measured mechanical quality factors are below \(4.0\times10^4\), the contact loss through the acrylic tape is not expected to limit the observed mechanical linewidths.

\section{Cavity simulation and mechanical mode area estimation}

Here we present the simulated electric-field profiles of the cavities at four representative frequencies. As shown in Fig. 2, from left to right, the four simulations correspond to the devices shown in Fig.~1(c), Fig.~2(a), Fig.~2(b), and Fig.~2(c) of the Main Text, respectively. The upper panels in Fig. 2(a) show the $x$-$y$ sectional views, illustrating the electric-field distribution inside and around the bulk LN wafer. The lower panels in Fig. 2(b) show the $y$-$z$ sectional views, illustrating the cavity mode volume and microwave confinement.

\begin{figure}[t]
\centering
\includegraphics[width=1\columnwidth]{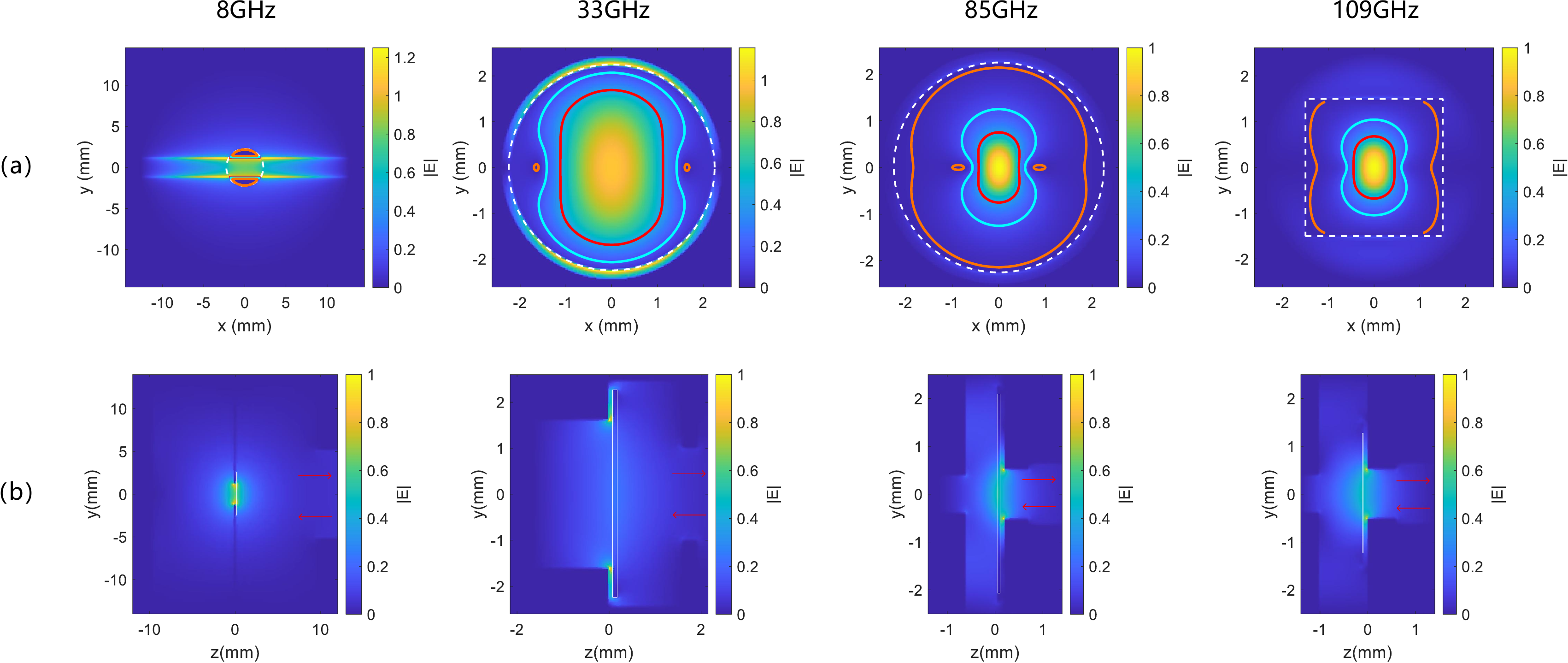}
\caption{Cavity mode simulations for four different devices operating at 8~GHz, 33~GHz, 85~GHz, and 109~GHz, respectively. (a) $x$-$y$ sectional views of the relative electric-field amplitude. The maximum electric-field amplitude inside the LN wafer is normalized to 1. The red, blue, and orange curves denote the $1/e$, $1/e^2$, and $1/e^4$ contour lines of the electric field, respectively, and the white dashed lines indicate the edge of the LN wafer. For the 8~GHz, 33~GHz, and 85~GHz devices, the LN wafers are circular disks with a diameter of 4.5~mm, while the wafer used for the 109~GHz device is a square with a side length of 3~mm. (b) $y$-$z$ sectional views of the relative electric-field amplitude. The white lines indicate the edge of the LN wafer.}
\label{Fig.7}
\end{figure}

Based on the discussion in Section~III, the transverse distribution of the mechanical mode amplitude is approximately proportional to the electric-field distribution. Therefore, Fig. 2(a) also provides an estimate of the transverse mechanical mode profile, which can be used to evaluate the mechanical mode area.

We estimate the mechanical mode area following the methods used in Refs.~\cite{ref29a, ref41a, ref47a}. In these references, the transverse mechanical mode profiles follow a Gaussian distribution, i.e., $u\propto e^{-r^2/r_0^2}$, and the mechanical mode area used for effective-mass estimation is defined as the area of a circle with radius $R=2r_0$. In our devices, however, the mechanical mode profile does not strictly follow a Gaussian distribution in the weak-field region, as indicated by the orange $1/e^4$ contour lines in Fig. 2(a). Therefore, we adopt a relatively conservative estimate and define the mechanical mode area as the area enclosed by the blue $1/e^2$ contour line. Based on our simulations, the mechanical mode areas of the four devices are estimated to be 10.2~mm$^2$, 11.1~mm$^2$, 3.3~mm$^2$, and 2.5~mm$^2$, respectively. These areas contain $99.8\%$, $97.0\%$, $94.4\%$, and $93.5\%$ of the total mechanical energy, respectively.

\section{Time-domain analysis through Fourier transform}

Here we numerically analyze the time-domain ringdown response of our cavity--phonon system from the measured frequency-domain reflection spectra. Since the piezoelectric cavity--phonon system studied here is a linear system, the complex frequency-domain response $S_{11}$ and the time-domain response can be converted to each other through a Fourier transform. For each center frequency of the probe pulse $f_{\rm probe}$, we calculate the time-domain ringdown response as
\begin{equation}
    a(t;f_{\rm probe}) =
    \int df
    W(f-f_{\rm probe})
    S_{11}(f)
    e^{i(f-f_{\rm probe})t},
\end{equation}
where $W(f-f_{\rm probe})$ is a window function centered at $f_{\rm probe}$. Here we use a sinc window function,
\begin{equation}
    W(f-f_{\rm probe})
    =
    {\rm sinc}\!\left(
    \frac{f-f_{\rm probe}}{{\rm BW}/2}
    \right),
    \qquad
    {\rm sinc}(x)=\frac{\sin(\pi x)}{\pi x},
\end{equation}
where we set the bandwidth ${\rm BW}=600~{\rm MHz}$. This window function accounts for the finite bandwidth of the probe pulse, and the finite bandwidth of the filtering and sampling rates of the measurement instrument, such as a lock-in amplifier, in an actual ringdown measurement.

As shown in Supplementary Fig.~3, we apply this procedure to the measured $S_{11}$ spectra shown in Fig.~3(a) of the Main Text. For each cavity tuning point, we set $f_{\rm probe}=f_a$ and obtain the reconstructed normalized ringdown color map shown in Supplementary Fig.~3(a). We then select seven traces at the resonant conditions $f_{\rm probe}=f_a\approx f_b$ and plot them in Supplementary Fig.~3(b).

To quantify the oscillatory ringdown, we average these seven normalized time-domain traces and fit the averaged response using a two-hybrid-mode model,
\begin{equation}
    |a_{\rm fit}(t)|
    =
    e^{-\Gamma_{\rm hybrid} t/2}
    \left|
    \cos\left(\frac{\Omega_{\rm splitting} t}{2}\right)
    \right|,
\end{equation}
where $\Gamma_{\rm hybrid}$ is the energy decay rate of the two hybrid modes and $\Omega_{\rm splitting}$ is their angular frequency splitting.

\begin{figure}[t]
\centering
\includegraphics[width=1\columnwidth]{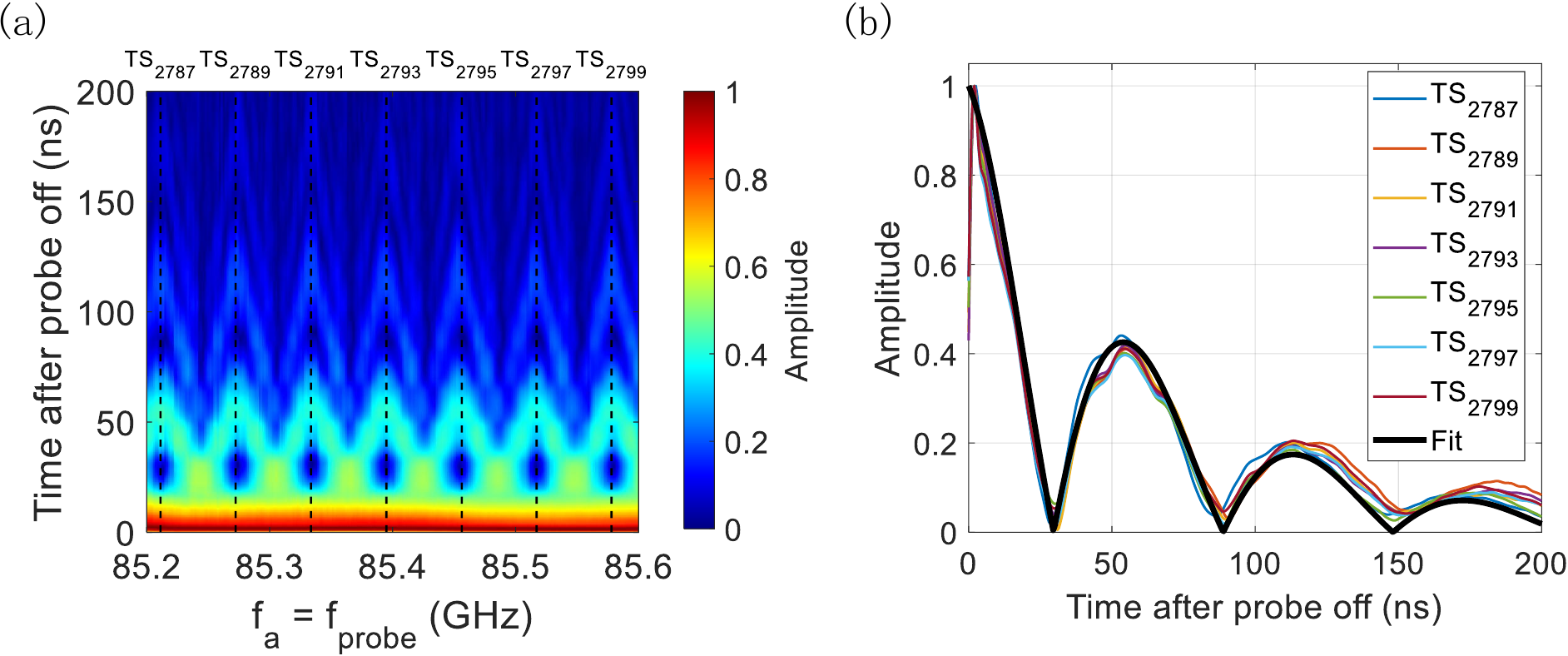}
\caption{Time-domain ringdown analysis. (a) Numerically calculated time-domain ringdown response obtained from the measured $S_{11}$ spectra shown in Fig.~3(a) of the Main Text. Clear oscillations are observed at the seven mechanical-mode frequencies, indicated by the black dashed lines. (b) Seven selected traces closest to the mechanical-mode frequencies in (a). The black curve shows the fitting for them.}
\label{Fig.8}
\end{figure}

The fit to this ringdown response gives a decay rate of $\Gamma_{\rm hybrid} = 2\pi\times4.80~{\rm MHz},$ and an oscillation rate of $\Omega_{\rm splitting} = 2\pi\times16.88~{\rm MHz}.$
These values agree well with the hybrid-mode linewidth $\frac{\Gamma_{a,\rm load}+\Gamma_b}{2}=2\pi\times5.04~{\rm MHz},$ and the coupling rate $2g = 2\pi\times17.08~{\rm MHz},$
obtained from the frequency-domain fitting, with relative differences within $5\%$. This agreement provides a further validation of the cavity--phonon coupling model and the frequency-domain fitting procedure used in this work.

\section{Power- and Temperature- dependence of the cavity loss.}

We characterize the device shown in Fig.~3(a) of the Main Text using power- and temperature-dependent $S_{11}$ measurements. As shown in Fig.~4, the input power is swept from $-50$ to $0$~dBm [Fig.~4(a)], with the transmission-line attenuation calibrated, while the temperature is swept from 3.65 to 8.05~K [Fig.~4(b)]. The corresponding fitting parameters are plotted in Figs.~4(c) and 4(d), respectively. In both cases, the cavity resonance shifts downward in frequency and its linewidth increases. These trends are attributed to the reduced Cooper-pair density and increased penetration depth at higher input powers or temperatures, which increase the effective mode volume and surface resistance of the superconducting cavity. Meanwhile, no significant increase in the mechanical decay rate $\Gamma_b$ is observed. This highlights the strong power and temperature tolerance of the mechanical resonator at cryogenic temperatures and indicates that the acrylic tape used in this work efficiently dissipates thermal energy from the bulk LN, thereby maintaining the optimal performance of the mechanical resonator even under mW-scale microwave heating.

\section{Dielectric loading and cavity intrinsic loss analysis}
Here we characterize the bare superconducting niobium cavity without the lithium niobate wafer. The simulated electric-field distribution is shown in Fig.~5(a). The simulation shows that, for our 85.6~GHz LN-loaded device with a cavity length of 0.8~mm, removing the LN wafer shifts the fundamental TE-mode resonance frequency upward to 185~GHz and substantially increases the mode volume. To bring this mode back into the W-band, we increased the cavity length to about 1.8~mm, reducing the resonance frequency to 93~GHz.
\begin{figure}[t]
    \centering
    \includegraphics[width=1.05\columnwidth]{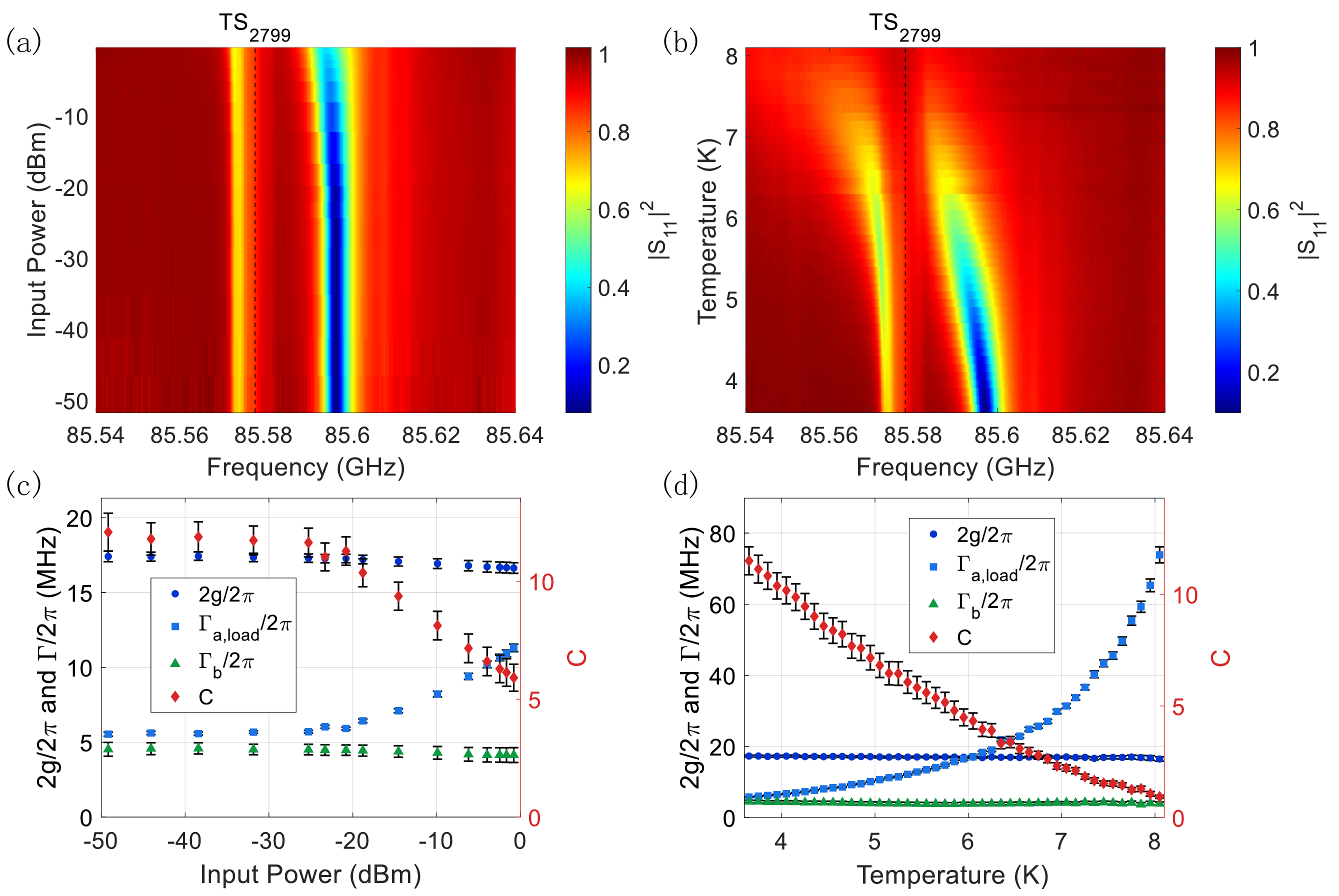}
    \caption{
    Power and temperature dependence. (a,b) Color maps of the reflection spectra of the device under varying input power and temperature, respectively. In (a), the input power is swept from $-50$ to $0$~dBm at a fixed temperature of 3.65~K, with the transmission-line attenuation calibrated to within $\pm 1$~dBm uncertainty. In (b), the temperature is swept from 3.65 to 8.05~K at a fixed input power of $-25$~dBm. (c,d) Extracted coupling rate $2g$, loaded cavity linewidth $\Gamma_{a,\rm load}$, mechanical linewidth $\Gamma_b$, and cooperativity $C$ for the corresponding powers and temperatures, respectively.
    }
\end{figure}

We measured the reflection spectra of the bare cavity over the temperature range from 4.25~K to 8.05~K, as shown in Fig.~5(b). A resonance frequency shift of 10.5~MHz is observed. Considering the cavity length of 1.8~mm, this shift corresponds to an effective penetration-depth change of approximately 100~nm from 4.25~K to 8.05~K, which is reasonable for niobium superconductors. In comparison, for the 85.6~GHz LN-loaded cavity, we observe a resonance frequency shift of 40~MHz, as shown in Fig.~4(b). This larger shift indicates that LN loading reduces the effective volume-to-surface ratio of the fundamental TE mode by approximately a factor of four. The extracted intrinsic loss rates $\Gamma_i$ of the bare cavity and the 85.6~GHz LN-loaded cavity at different temperatures are plotted in Fig.~5(c). The intrinsic loss rates of both devices decrease exponentially with decreasing temperature and this trend does not show obvious saturation at the lowest temperature tested. It indicates that microwave absorption associated with the niobium surface resistance is one of the dominant factors limiting the quality factor at liquid-helium temperatures. Meanwhile, the loss-rate ratio between the LN-loaded and bare cavities is approximately 4--6 in this temperature range, which is in reasonable agreement with the volume-to-surface ratio estimated from the resonance frequency shift.

\begin{figure}[t]
    \centering
    \includegraphics[width=1.05\columnwidth]{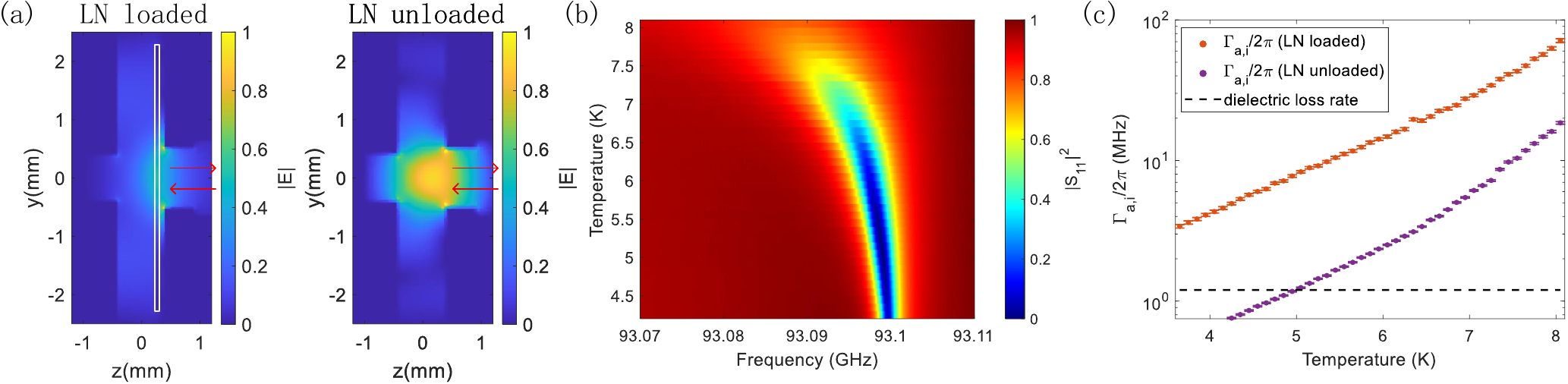}
    \caption{
    (a) Microwave simulation for the 85~GHz LN-loaded device before and after removing the LN wafer. The resonance frequency increases to 185~GHz when removing the wafer. (b) Color map of the reflection spectra for the temperature dependence measurement of the bare cavity. (c) The loss rate for the bare cavity (purple points, $f_a=93.1$~GHz) and LN-loaded cavity (orange points, $f_a=85.6$~GHz), together with the dielectric loss rate $2\pi\times1.2$~MHz (black dashed line).
    }
\end{figure}

Another important loss channel for the LN-loaded cavity is the dielectric loss. According to Ref.~\cite{ref50a}, the loss tangent of lithium niobate is $1.73\times10^{-5}$ at 7.2~GHz and 50~mK. Using this value as an estimate, and since about $80\%$ of the energy is in the LN wafer based on the simulation, we estimate the dielectric loss rate to be
$2\pi\times1.2$~MHz.
This estimate is plotted as the black dashed line in Fig.~5(c). Note that there should be an uncertainty because the dielectric loss tangent of lithium niobate at 85~GHz and 3.65~K may differ from the reported value at 7.2~GHz and 50~mK. At 3.65~K, the intrinsic loss rate of the LN-loaded cavity is $2\pi\times3.40$~MHz ($Q_{a,i}=25,100$), indicating that the dielectric loss can account for about $35\%$ of the total energy decay and may ultimately limit the cavity quality factor to approximately $70,000$ below 2~K. In contrast, the bare cavity is free from dielectric loss and reaches an intrinsic loss rate of $2\pi\times0.76$~MHz ($Q_{a,i}=122,900$) at 4.25~K owing to the high volume-to-surface ratio of the cavity mode.

Furthermore, since the microwave cavities in this work are assembled from two machined niobium halves, insufficient flatness after machining or slight misalignment between the two halves can leave a small gap along the cavity sidewall, potentially limiting the cavity quality factor to approximately $10^4$. This issue is eliminated by using new drill bits and end mills for each machining process and pressing the two halves together using an arbor press after machining to ensure a flat, well-matched interface. As confirmed by the bare-cavity measurements, the residual loss from this effect is much smaller than $2\pi\times1.2$~MHz via this improved machining process. 

\begin{figure}[t]
    \centering
    \includegraphics[width=1\columnwidth]{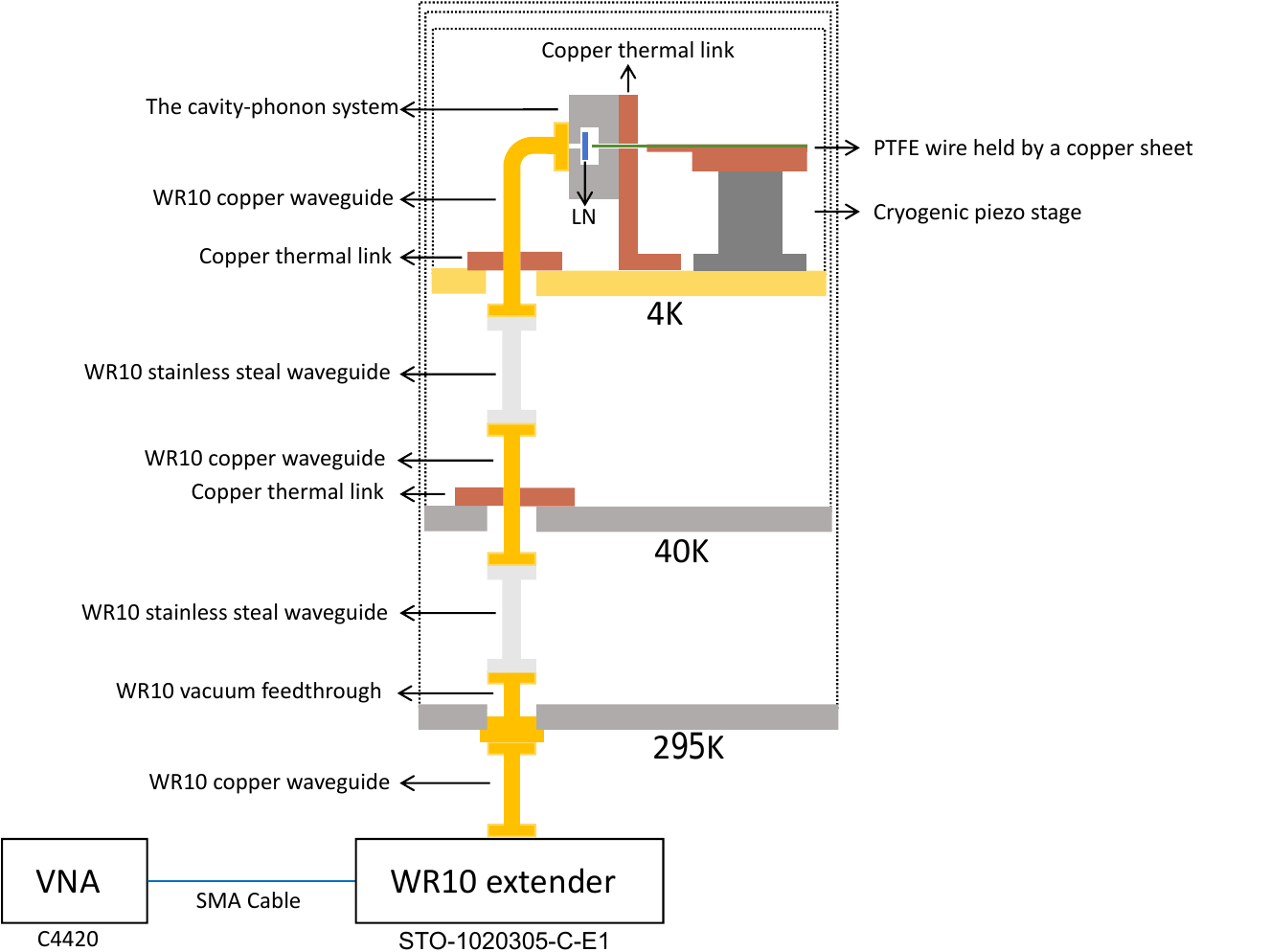}
    \caption{
    Schematic of the experimental setup. The measurement chain and the cavity-tuning apparatus are illustrated.
    }
\end{figure}

\section{Experimental Setup}

Figure~6 shows the schematic of the experimental setup. The reflection spectrum is measured using a commercial 0--20~GHz vector network analyzer (VNA) together with a commercial WR10 frequency extender. The fridge in this work is a closed-cycle cryostat. The microwave transmission line from room temperature to the 4~K stage is constructed using WR10 waveguides and consists of both copper and stainless-steel sections. The stainless-steel sections reduce the thermal load between different temperature stages, while the copper sections provide efficient thermalization of the waveguides to the corresponding cryostat stages. The cavity resonance frequency is tuned using a PTFE dielectric wire inserted into the cavity. The insertion depth of the PTFE wire is controlled by the horizontal axis of a cryogenic piezoelectric stage.

\bibliographystyle{apsrev4-1}
\bibliography{main}